\documentclass[conference,twocolumn]{IEEEtran}

\usepackage{amsmath,graphicx}
\usepackage{color}
\usepackage{graphicx}
\usepackage{epstopdf}
\usepackage{amsmath}
\usepackage{amssymb}
\usepackage{mathrsfs}
\usepackage{hyperref}
\usepackage{tikz}
\usepackage{bm}
\usepackage[english]{babel}
\usepackage{cite}
\usepackage{rotfloat}
\usepackage{mathtools}
\usepackage[font=normalsize,labelfont=bf]{caption}
\usepackage{amsmath}
\usepackage{makecell}
\usepackage{multirow}
\usepackage{booktabs}
\usepackage{colortbl}
\usepackage{multirow}
\usepackage{hhline}
\usepackage{stfloats}
\usepackage{cuted} 
\usepackage{multicol}
\usepackage{bbm}
\usepackage{cases}
\graphicspath{ {Figures/} }
\newcommand{\T}{{\scriptscriptstyle\mathsf{T}}}
\renewcommand{\H}{{\scriptscriptstyle\mathsf{H}}}

\newsavebox{\foobox}

\definecolor{kugray5}{RGB}{224,224,224}

\usepackage[normalem]{ulem}
\newcommand\rsout{\bgroup\markoverwith
	{\textcolor{red}{\rule[0.5ex]{2pt}{0.8pt}}}\ULon}



\makeatletter
\newcommand{\ALOOP}[1]{\ALC@it\algorithmicloop\ #1%
	\begin{ALC@loop}}
	\newcommand{\ENDALOOP}{\end{ALC@loop}\ALC@it\algorithmicendloop}

\makeatother

\usepackage{etoolbox}
\let\mybibitem\bibitem
\renewcommand{\bibitem}[1]{%
	\ifstrequal{#1}{nature}
	{\color{blue}\mybibitem{#1}}
	{\color{black}\mybibitem{#1}}%
}

\graphicspath{ {Figures/} }

\DeclareCaptionLabelSeparator{periodspace}{.\quad}

\addto\captionsenglish{}
\allowdisplaybreaks

\usepackage{setspace}

\newcommand{\mW}{{\mathbf{W}}}

\newcommand{\mY}{{\mathbf{Y}}}

\newcommand{\ve}{{\mathbf{e}}} 
\newcommand{\vs}{{\mathbf{s}}}
\newcommand{\vx}{{\mathbf{x}}}
\newcommand{\vy}{{\mathbf{y}}}

\newcommand{\vu}{{\mathbf{u}}}
 
\newcommand{\vh}{{\mathbf{h}}}

\newcommand{\va}{{\mathbf{a}}}

\newcommand{\vg}{{\mathbf{g}}}

\newcommand{\vphi}{\boldsymbol{\varphi}}

\def\b0{{\pmb{0}}}

\usepackage{mdframed} 

\usepackage{caption} 
\usepackage{subcaption} 
\usepackage{appendix} 
\usepackage[ruled,vlined]{algorithm2e} 

\begin{document}
\title{Deep Reinforcement Learning-Based Dynamic Resource Allocation in Cell-Free Massive MIMO}

\author{
    \IEEEauthorblockN{Phuong~Nam~Tran\IEEEauthorrefmark{1}, Nhan~Thanh~Nguyen\IEEEauthorrefmark{1}, Hien~Quoc~Ngo\IEEEauthorrefmark{2}, Markku~Juntti\IEEEauthorrefmark{1},}
    \IEEEauthorblockA{
    \IEEEauthorrefmark{1}Centre for Wireless Communications (CWC), University of Oulu, Finland, \\
    \IEEEauthorrefmark{2}School of Electronics, Electrical Engineering and Computer Science, Queen’s University Belfast, UK
    \\\{phuong.tran, nhan.nguyen, markku.juntti\}@oulu.fi, hien.ngo@qub.ac.uk
    }
    }
	
\maketitle
	
\begin{abstract}
     In this paper, we consider power allocation and antenna activation of cell-free massive multiple-input multiple-output (CFmMIMO) systems. We first derive closed-form expressions for the system spectral efficiency (SE) and energy efficiency (EE) as functions of the power allocation coefficients and the number of active antennas at the access points (APs). Then, we aim to enhance the EE through jointly optimizing antenna activation and power control. This task leads to a non-convex and mixed-integer design problem with high-dimensional design variables. 
     To address this, we propose a novel DRL-based framework, in which the agent learns to map large-scale fading coefficients to AP activation ratio, antenna coefficient, and power coefficient.
     These coefficients are then employed to determine the number of active antennas per AP and the power factors assigned to users based on closed-form expressions.
     By optimizing these parameters instead of directly controlling antenna selection and power allocation, the proposed method transforms the intractable optimization into a low-dimensional learning task.
     Our extensive simulations demonstrate the efficiency and scalability of the proposed scheme. Specifically, in a CFmMIMO system with 40 APs and 20 users, it achieves a 50\% EE improvement and 3350 times run time reduction compared to the conventional sequential convex approximation method.
     \end{abstract}

\begin{IEEEkeywords}
Cell-free massive MIMO, energy efficiency, deep reinforcement learning, antenna selection, power control.
\end{IEEEkeywords}

\IEEEpeerreviewmaketitle

\section{Introduction}
Recently, cell-free massive multiple-input multiple-output (CFmMIMO) has emerged as a promising technology for next generation wireless systems. In CFmMIMO networks, a large number of access points (APs) are deployed in a large area to coherently serve users and offer uniformly great quality of service (QoS) to all users with simple signal processing \cite{ngo2017cell, ngo2017total}.
However, dense AP deployment in CFmMIMO imposes significant energy demands, making energy efficiency (EE) a key priority for the system design \cite{ngo2024ultradense}. 

Maximizing EE for CFmMIMO systems often entails a joint optimization of power allocation together with the dynamic selection of APs and/or antennas, both involving high-dimensional variables due to the large size of CFmMIMO networks \cite{ngo2017total, nguyen2022hybrid}.
Conventional optimization approaches to EE maximization, such as successive convex approximation (SCA) \cite{ngo2017total} and accelerated projected gradient \cite{mai2022energy} often offer good performance. However, they are generally iterative in nature, requiring high computational complexity and long run time. Thus, these methods have limited practicality for real-time operation of large-size systems as CFmMIMO \cite{ngo2017total, alonzo2019energy, bashar2019energy, mai2022energy}.

Deep reinforcement learning (DRL) has emerged as a promising paradigm for real-time operations \cite{wang2024green}. However, existing DRL frameworks for EE design and optimization in CFmMIMO systems face critical scalability issues.
Specifically, DRL approaches for power control \cite{luo2022downlink} and user association \cite{ghiasi2022energy, mendoza2023user} often require action spaces whose dimensionality grows with the network size.  
Similarly, multi-agent DRL schemes \cite{liu2024joint} typically scale the number of agents  with the number of users, further amplifying the scalability issue. Therefore, these DRL solutions can be unsuitable for large-scale CFmMIMO networks.
Conversely, scalable DRL designs that randomly deactivate APs according to deployment density oversimplify the problem and can result in suboptimal performance \cite{topal2023drl}.
Therefore, developing a DRL framework that is simultaneously scalable and capable of jointly optimizing power allocation, AP deactivation, and per-AP antenna selection to maximize EE remains a critical open challenge.

In this work, we aim to enhance the EE of CFmMIMO systems. We first derive closed-form expressions for the system spectral efficiency (SE) and EE. Then, we focus on optimizing antenna activation and power control for EE maximization. 
To ensure scalability, we propose a DRL-based framework in which the agent maps the network’s large-scale fading coefficients to AP activation ratio, antenna allocation, and a power allocation coefficients, which allow to compute the number of antennas and power factors at APs using closed-form expressions. 
This design transforms the high-dimensional and intractable problem into a compact, fixed-size action space, enabling scalable and low-complexity DRL-based optimization.
Simulations demonstrate that the proposed approach can enhance EE by 50\% and reduce runtime by approximately 3350 times in a CFmMIMO system with $40$ APs and $20$ users, outperforming the conventional SCA method.

\section{System Model}
\label{sec:system_model}
\begin{figure}[t!]
    \centering
    \includegraphics[width=0.9\columnwidth]{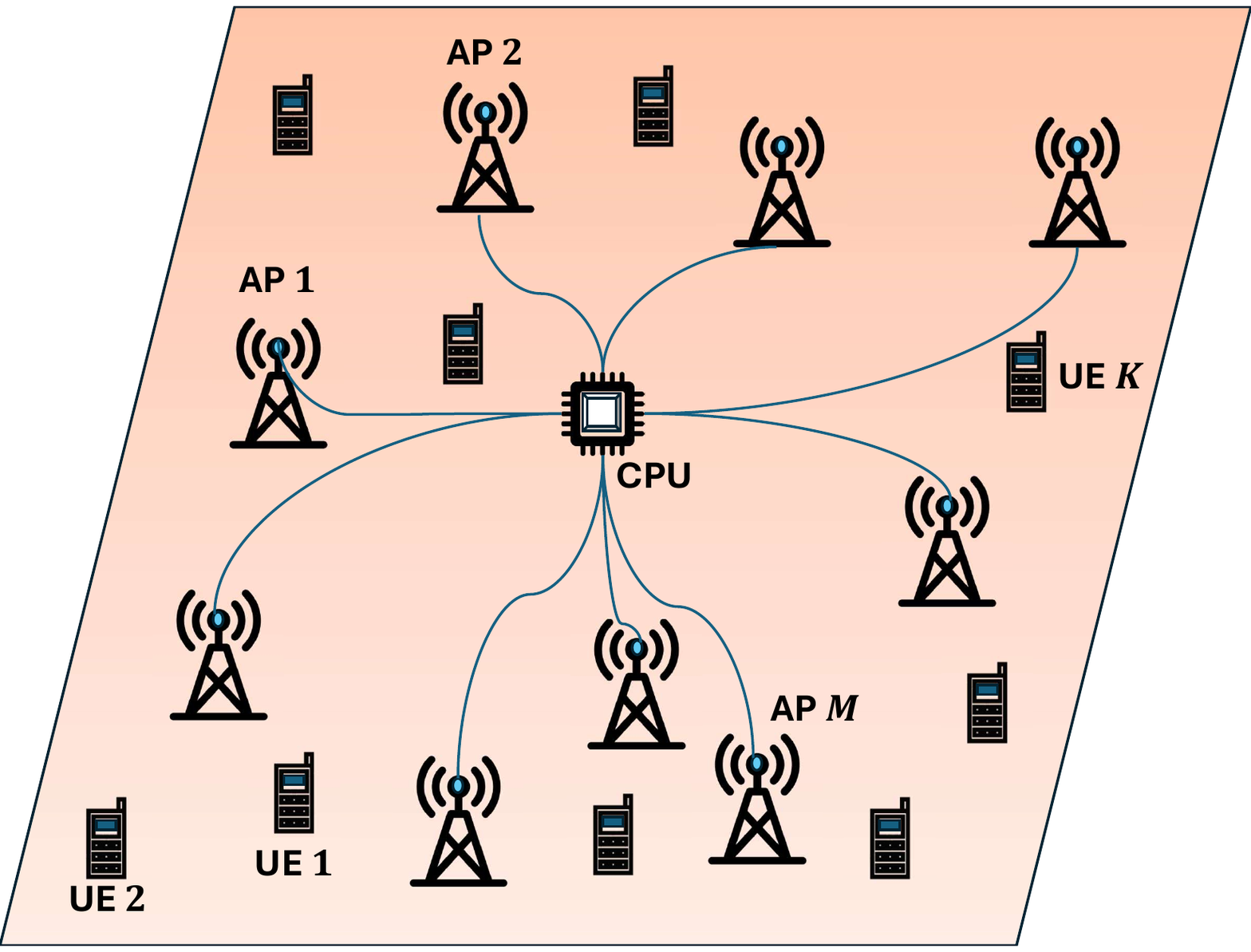} 
    \caption{Cell-free massive MIMO system.}
    \label{fig:system_model}
\end{figure}

We consider a CFmMIMO system in which $M$ APs, each equipped with $N$ antennas, serve $K$ single‑antenna users over the same time‑frequency resources, with $MN \gg K$ \cite{ngo2017total}. All APs connect to a central processing unit (CPU) via a backhaul network, as illustrated in Fig.~\ref{fig:system_model}. 
We consider the scenario that at each AP, only a subset of antennas is activated to reduce the power consumption. Let $N_m \leq N$ be the number of activated antennas at AP $m$.
 
\subsection{Channel Estimation}
During the uplink training phase, all $K$ users simultaneously transmit their assigned pilot sequences. Let $\tau_\mathsf{c}$ and $\tau_\mathsf{p}$ respectively denote the lengths of each coherent interval and pilot sequence $\vphi_k$ (in samples). Let $\sqrt{\tau_\mathsf{p}}\vphi_k \in \mathbb{C}^{\tau_\mathsf{p} \times 1}$ denote the pilot sequence of user $k$, where $\|\vphi_k\|^2 = 1$.
For each AP $m$, we define a binary antenna activation vector $\mathbf{s}_m = [s_{m1}, \ldots, s_{mN}]$, where $s_{mn} = 1$ indicates that the $n$-th antenna is active, while $s_{mn} = 0$ indicates that it is deactivated. The number of active antennas at AP $m$ is $\|\vs_m\|_1 = N_m$.
 
The signal received at AP $m$ is given by
\begin{equation}
    \mY_{\mathsf{p},m} = \sqrt{\tau_\mathsf{p} \rho_\mathsf{p}} \sum_{k=1}^{K} (\vs_m \odot \vg_{mk}) \vphi_k^\H + (\vs_m \odot \mW_{\mathsf{p},m}),
\end{equation}
where $\rho_\mathsf{p}$ denotes the normalized transmit signal-to-noise ratio (SNR) per pilot symbol, $\mW_{\mathsf{p},m} \in \mathbb{C}^{N \times \tau_\mathsf{p}}$ is an additive white Gaussian noise matrix, and $\odot$ represents the element-wise (Hadamard) product. The channel vector $\vg_{mk} \in \mathbb{C}^{N \times 1}$ is modeled as $\vg_{mk} = \beta_{mk}^{1/2} \vh_{mk}$, where $\beta_{mk}$ is the large-scale fading coefficient, and $\vh_{mk}$ contains independent and identically distributed (i.i.d.) $\mathcal{CN}(0, 1)$ entries of small-scale fading channel.

We define the effective channel between user $k$ and the active antennas of AP $m$ as $\vg_{mk}^{\text{A}} \triangleq \vs_m \odot \vg_{mk}$.
The effective channel $\vg_{mk}^{\text{A}}$ therefore contains only the components corresponding to the active antennas, which are involved in channel estimation and data transmission.
To estimate $\vg_{mk}^{\text{A}}$, $\mY_{\mathsf{p},m}$ is projected onto $\vphi_k$ to yield $\tilde{\vy}_{\mathsf{p},mk} \triangleq \mY_{\mathsf{p},m}\vphi_k$.
The minimum mean-square error (MMSE) estimate of the channel $\vg_{mk}$ is given by \cite{kay1993fundamentals}
\begin{equation}
    \hat{\vg}^{\text{A}}_{mk} = \frac{\sqrt{\tau_\mathsf{p}\rho_\mathsf{p}}\beta_{mk}}{\tau_\mathsf{p}\rho_\mathsf{p} \sum_{k'=1}^{K} \beta_{mk'}|\vphi_{k'}^\H\vphi_k|^2+1} \tilde{\vy}_{\mathsf{p},mk}.
\end{equation}
Let $\hat{g}^{\text{A}}_{mkn}$ denote the $n$-th element of $\hat{\mathbf{g}}^{\text{A}}_{mk}$. Due to antenna selection, we have $\hat{g}^{\text{A}}_{mkn} = 0$ if $s_{mn} = 0$ , while for $s_{mn} = 1$, the mean square of $\hat{g}^{\text{A}}_{mkn}$ is given as \cite{ngo2017total}
\begin{align} \label{eq_gamma}
    \mathbb{E}\{|\hat{g}^{\text{A}}_{mkn}|^2\} &= \frac{\tau_\mathsf{p} \rho_\mathsf{p} \beta_{mk}^2}{\tau_\mathsf{p} \rho_\mathsf{p} \sum_{k'=1}^{K} \beta_{mk'} |\vphi_{k'}^\H \vphi_k|^2 + 1} \triangleq \gamma_{mk}. 
\end{align}
The estimation error associated with $\hat{\vg}_{mk}$, denoted by $\tilde{\ve}_{mk}$, is statistically independent of the estimate $\hat{\vg}_{mk}$, and its elements are i.i.d. $\mathcal{CN}(0, \beta_{mk}-\gamma_{mk})$ random variables.

\subsection{Downlink Data Transmission}
The APs use the channel estimates to perform conjugate beamforming, i.e., the beamforming vector at AP $m$ towards user $k$ is set to $\hat{\vg}_{mk}^*$. Let $q_k$ be the data symbol intended for user $k$, with $\mathbb{E}[|q_k|^2] = 1$. The signal transmitted from AP $m$ is given as
\begin{equation}
\label{eq:transmit_signal}
    \vx_m = \sqrt{\rho_\mathsf{d}} \sum_{k=1}^{K} \sqrt{\eta_{mk}} \hat{\vg}_{mk}^* q_k,
\end{equation}
where $\rho_\mathsf{d}$ is the maximum normalized transmit power, and $\eta_{mk}$ is the power control coefficient. The total normalized power from AP $m$ is $\mathbb{E}\{\|\vx_m\|^2\} = \rho_\mathsf{d} N_m \sum_{k=1}^{K} \eta_{mk} \gamma_{mk}$. The coefficients are selected to satisfy the per-AP power constraint $ \sum_{k=1}^{K} \eta_{mk} \gamma_{mk} \le \frac{1}{N_m}, \forall m$.
The signal received at user $k$ is given by
\begin{equation}
\label{eq:received_signal}
    r_k = \sum_{m=1}^{M} (\vg_{mk}^{\text{A}})^\T \vx_m + w_k,
\end{equation}
where $w_k \sim \mathcal{CN}(0, 1)$ is the additive noise at user $k$.

\section{Problem Formulation}
\label{sec:problem_formulation}
In this section, we formulate the EE maximization problem. We begin by deriving the closed-form SE, followed by the power consumption model. These components are then integrated to derive the EE and formulate the final optimization problem.

\subsection{Spectral Efficiency}
Each user detects its signal based on statistical channel knowledge, eliminating the need for downlink pilots \cite{ngo2017cell, marzetta2016fundamentals}. 
The received signal $r_k$ in \eqref{eq:received_signal} can be decomposed as
\begin{align}
    r_k =\;& 
    \mathrm{DS}_k q_k + \mathrm{BU}_k q_k + \sum_{j \neq k}^{K} \mathrm{UI}_{kj}  q_j + w_k.
\end{align}
Here, $\mathrm{DS}_k \triangleq \sqrt{\rho_\mathsf{d}} \mathbb{E} \{\sum_{m=1}^{M} \sqrt{\eta_{mk}} \vu_{mkk} \}$, $\mathrm{BU}_k \triangleq \sqrt{\rho_\mathsf{d}} (\sum_{m=1}^{M} \sqrt{\eta_{mk}} \vu_{mkk} - \mathbb{E} \{\sum_{m=1}^{M} \sqrt{\eta_{mk}} \vu_{mkk} \} )$, and $\mathrm{UI}_{kj} \triangleq \sqrt{\rho_\mathsf{d}} \sum_{m=1}^{M} \sqrt{\eta_{mj}} \vu_{mkj}$, with $\vu_{mkj} \triangleq (\vg_{mk} \odot \vs_m)^\T \hat{\vg}_{mj}^*$, represent the desired signal, beamforming uncertainty, and inter-user interference power, respectively.  
Accordingly, the SE for user $k$ is then given by \cite{marzetta2016fundamentals}
\begin{equation}
\small
\label{eq:original_Sek}
S_{\mathrm{e},k} = \frac{\tau_\mathsf{c} - \tau_\mathsf{p}}{\tau_\mathsf{c}} 
\log_2 \!\left( 1 + \frac{|\mathrm{DS}_k|^2}{\mathbb{E}\{|\mathrm{BU}_k|^2\} + \sum_{j \neq k}^{K} \mathbb{E}\{|\mathrm{UI}_{kj}|^2\} + 1} \right).
\end{equation}
Then, the total SE is obtained as $S_{\mathrm{e}} = \sum_{k=1}^{K} S_{\mathrm{e},k}$. We present the closed-form expression for $S_{\mathrm{e},k}$ in \eqref{eq:spectral_efficiency}, shown at the top of next page, and omit the detailed proof here due to the limited space. It is observed from \eqref{eq:spectral_efficiency} that the SE depends on the numbers of active antennas at the APs, i.e., $\{N_m\}_{m=1}^M$, rather than on the specific indices of the active antennas.

\begin{figure*}[!t] 
\centering
\begin{equation}
\label{eq:spectral_efficiency}
S_{\mathrm{e},k} \triangleq 
    \frac{\tau_\mathsf{c} -\tau_\mathsf{p}}{\tau_\mathsf{c}}
    \log_2 \left( 1+ \frac
    {\rho_\mathsf{d} \left( \sum_{m=1}^{M} \sqrt{\eta_{mk}} N_m \gamma_{mk} \right)^2}
    {
    \rho_\mathsf{d} \sum_{j \neq k}^{K} \left( \sum_{m=1}^{M} N_m \sqrt{\eta_{mj}} \gamma_{mj} \frac{\beta_{mk}}{\beta_{mj}} \right)^2
    + \rho_\mathsf{d} \sum_{m=1}^{M} \beta_{mk} N_m \left( \sum_{j=1}^{K} \eta_{mj} \gamma_{mj} \right)
    + 1 
    }\right)
\end{equation}
\hrulefill
\end{figure*}

\subsection{Power Consumption}
The total power consumption of the network is the sum of the power consumed at each AP and by the associated backhaul links, given by
\begin{align}
\label{eq:p_total}
P_{\text{total}} = &\sum_{m=1}^{M} 
\Big( 
\underbrace{\tfrac{1}{\alpha_m} \rho_\mathsf{d} N_0 \!\Big( N_m \sum_{k=1}^{K} \eta_{mk} \gamma_{mk} \Big) 
+ N_m P_{\text{tc},m}}_{\triangleq P_m} \nonumber \\
&+ \underbrace{P_{0,m} + B \cdot S_{\mathrm{e}} \cdot P_{\text{bt},m}}_{\triangleq P_{\text{bh},m}}
\Big).
\end{align}
Here, $P_m$ is the power at AP $m$, consisting of the amplified transmit power which depends on the thermal noise power $N_0$, scaled by the power amplifier efficiency $\alpha_m \in (0,1]$, 
and the static circuit power with $P_{\text{tc},m}$ denoting the per-antenna circuit power.  
The backhaul power $P_{\text{bh},m}$ includes the fixed power consumption of each backhaul $P_{0,m}$ and a traffic-dependent term, where $B$ is the system bandwidth, 
$S_{\mathrm{e}}$ is the sum SE, and $P_{\text{bt},m}$ is the power per bit/s.

\subsection{EE Maximization Problem}
The network EE, measured in bits per Joule, is defined as the ratio of the total throughput to the total power consumption, i.e., $E_{\mathrm{e}} = \frac{B \cdot S_{\mathrm{e}}}{P_{\text{total}}}$. Since both the total power consumption and the SE depend only on the numbers of active antennas at the APs, i.e., $\{N_m\}$, the optimization of antenna activation should be exploited. Therefore, the joint optimization of antenna activation and power control in the considered CFmMIMO system can be formulated as
\begin{subequations}\label{eq:P1}
    \begin{align}
        (\mathrm{P1}):\ \underset{\{\eta_{mk}\}, \{N_m\}}{\text{maximize}} \quad &  E_{\mathrm{e}}   \label{eq:opt_EE} \\
        \text{subject to} \quad 
        & S_{\mathrm{e},k} \geq S_{\mathrm{ok}}, \quad \forall k \label{eq:cons_SE} \\
        & \sum_{k=1}^K \eta_{mk} \gamma_{mk} \leq \frac{1}{N_m}, \quad \forall m \label{eq:cons_Ptx} \\
        & N_m \in \{0, 1, \ldots, N\} , \quad \forall m \label{eq:cons_n_m}
    \end{align}
\end{subequations}
where \eqref{eq:cons_SE} ensures a minimum SE for each user, \eqref{eq:cons_Ptx} limits the total transmit power across all APs, and \eqref{eq:cons_n_m} imposes constraints on the number of active antennas at each AP, i.e., $\{N_m\}$. The formulated resource allocation problem is a fractional non-convex and mixed-integer program, which is highly challenging and cannot be solved directly by conventional optimization solvers. Furthermore, the large number of APs necessitates a scalable solution, which will be presented next.

\section{DRL for Resource Allocation}
\label{sec:proposed_framework}
We aim to develop an efficient DRL framework to solve \eqref{eq:P1}. However, a direct application of DRL is challenging and  impractical due to the prohibitively large action space, which consists of $N^M$ possibilities for antenna configurations and $MK$ power coefficients. To address this, we propose a DRL framework built on heuristic AP activation, antenna selection, and power allocation strategies. In this framework, instead of directly controlling the numbers of antennas $\{N_m\}$ and power coefficients $\{\eta_{mk}\}$, the DRL agent outputs the AP activation ratio $\zeta$, the antenna allocation coefficient $\kappa$, and the power allocation coefficient $\nu$, which are then used to compute $\{N_m, \eta_{mk}\}$ with closed forms. The AP activation, antenna selection and power allocation methods are presented next.

\subsection{Resource Allocation}
\label{sec:resource_allocation}
The key ideas for AP activation, antenna selection, and power allocation are outlined below.

\textbf{AP Activation:} We first assume equal power and antenna allocation across all APs. For AP selection, we propose choosing the APs with the largest average large-scale fading coefficients across all users. Specifically, for AP $m$, this metric is defined as
\begin{equation}
    \label{eq:importance_score}
    I_{\mathrm{AP},m} = \frac{1}{K} \sum_{k=1}^{K} \beta_{mk}.
\end{equation}
An AP with a larger $I_{\mathrm{AP},m}$ generally contributes more to the system’s EE and should therefore be selected for transmission. We introduce a parameter $\zeta$ as the threshold for AP selection, where the top $\zeta\%$ of APs with the largest $I_{\mathrm{AP},m}$ values are selected and form the set of active APs $\mathcal{A}_{\mathsf{active}}$. A smaller $\zeta$ reduces power consumption but may degrade both the SE and EE performances due to the limited number of active APs. Hence, we treat $\zeta$ as an optimization variable and optimize it using DRL, as detailed in the next section.

\textbf{Antenna Selection:} Given the active AP set $\mathcal{A}_{\mathsf{active}}$ obtained in the previous step, we propose to allocate more antennas to APs with larger $I_{\mathrm{AP},m}$, $m \in \mathcal{A}_{\mathsf{active}}$. Specifically, the number of active antennas at AP $m$ is determined as
\begin{equation}
    \label{eq:antennas_allocation}
    N_m = \lfloor 1 + (N-1) \cdot w_m \rfloor,
\end{equation}
where $\lfloor \cdot \rfloor$ denotes the floor operation, and
\begin{equation}
    \label{eq:activation_weight}
    w_m = \frac{I_{\mathrm{AP},m}^{\kappa}}{\max_{j \in \mathcal{A}_{\mathsf{active}}} I_{\mathrm{AP},j}^{\kappa}},
\end{equation}
with $\kappa \ge 0$ being a parameter that controls the antenna distribution across active APs. From \eqref{eq:antennas_allocation} and \eqref{eq:activation_weight}, it is observed that a larger $\kappa$ results in more antennas being assigned to APs with stronger large-scale fading coefficients, whereas a smaller $\kappa$ yields a more uniform allocation. By tuning $\kappa$, the network balances the SE gains from using more antennas against the corresponding increase in circuit power, thereby enhancing overall EE.

\textbf{Power allocation:} We adopt the fractional power allocation method \cite{demir2021foundations}, which assigns more power to users with larger values of $\gamma_{mk}$, $m \in \mathcal{A}_\mathsf{active}$, where  we recall that $\gamma_{mk}$ denotes the estimated channel gain as defined in \eqref{eq_gamma}. The power allocated by AP $m$ to user $k$ is then obtained as 
\begin{equation}
    \eta_{mk} = \frac{1}{N_m} \cdot \frac{\gamma_{mk}^{\nu-1}}{\sum_{j=1}^{K} \gamma_{mj}^{\nu}},
    \label{eq:power_allocation}
\end{equation}
which leads to $\sum_{k=1}^K \eta_{mk} \gamma_{mk} = \frac{1}{N_m}, \forall m$, satisfying constraint \eqref{eq:cons_Ptx}. In \eqref{eq:power_allocation}, $\nu \geq 0$ is an tunning parameter controlling the distribution of power among users. Larger $\nu$ leads to allocating more power to users with stronger $\gamma_{mk}$, thereby improving their SE and potentially enhancing overall EE. However, this increases the likelihood of violating the QoS constraint in \eqref{eq:cons_SE} for users with weaker channels. Conversely, smaller $\nu$ yield a more uniform allocation, but limit the power assigned to users with favorable channel conditions, potentially reducing SE and EE.

\subsection{DRL Agent Formulation}
\label{sec:rl_formulation}

With the resource allocation mechanisms described above, antenna selection and power allocation for EE maximization reduce to the joint optimization of $\{\zeta, \kappa, \nu\}$. We formulate this as a reinforcement learning (RL) task \cite{sutton1998reinforcement}, where a RL agent adaptively adjusts these parameters based on the observed network states. It is worth noting that all of these parameters depend only on the large-scale fading coefficients $\beta_{mk}$. Consequently, the stage remains unchanged within one large-scale fading interval. Let $t \in \{1,2,\ldots,T\}$ index the large-scale intervals, hereafter referred to as ``time slots'' for simplicity. 

In the proposed RL framework, the state, action, and reward at time slot $t$, are defined as follows:
\begin{itemize}
    \item \textbf{State:} The agent observes state $\vs_t = \boldsymbol{\beta}_t$, which collects the large-scale fading coefficients $\beta_{mk,t}$.
    \item \textbf{Action:} The agent outputs action $\va_t = [\zeta_t, \kappa_t, \nu_t]$, representing the resource allocation coefficients. This implies that the agent learns to map large-scale fading information to dynamic allocation decisions.
    \item \textbf{Reward:} Based on \eqref{eq:P1}, the reward encourages EE maximization while penalizing QoS violations, and is defined as
    \begin{equation}
        r_t(\vs_t, \va_t) = E_{\mathrm{e}} - \xi_{\mathrm{pen}} \sum_{k=1}^{K} \max\!\left(0, S_{\mathrm{ok}} - S_{\mathrm{e},k}\right),
        \label{eq:reward_function}
    \end{equation}
    where $\xi_{\mathrm{pen}} > 0$ is a penalty coefficient ensuring that the QoS constraint in \eqref{eq:cons_SE} is satisfied.
\end{itemize}

The RL objective is to learn a policy $\pi_\theta$ that maps states to actions to maximize the expected long-term reward:  
\begin{equation}
    (\mathrm{P2}): \quad \max_{\pi_\theta} \; \mathbb{E}\!\left[\sum_{t=0}^{T} \vartheta^t r_t \right],
    \label{eq:P2}
\end{equation}
where $T$ is the number of time slots, $0 \leq \vartheta \leq 1$ is the discount factor, and the expectation is taken over trajectories induced by $\pi_\theta$. This formulation serves as the basis for training the DRL agent to optimize resource allocation for EE maximization under QoS constraints.

\subsection{Proximal Policy Optimization}
To solve the RL problem in \eqref{eq:P2}, we employ proximal policy optimization (PPO) \cite{schulman2017proximal}, which effectively handles continuous variables and ensures stable learning by restricting policy updates to small, controlled steps. In PPO, the agent’s policy $\pi_\theta(\va_t|\vs_t)$ and value function $V_\phi(\vs_t)$ are parameterized by neural networks with parameters $\theta$ and $\phi$, respectively. The policy network outputs the mean and variance of a continuous action distribution from which $\va_t$ is sampled, while the value network estimates the expected return $V_\phi(\vs_t)$ to guide policy updates. The advantage function, estimated via generalized advantage estimation (GAE) \cite{schulman2017proximal}, is given by
\begin{equation}
\hat{A}_t = \sum_{l=0}^{T-t-1} (\vartheta \lambda)^l \delta_{t+l}, \quad
\delta_t = r_t + \vartheta V_\phi(\vs_{t+1}) - V_\phi(\vs_t),
\label{eq:advantage_func}
\end{equation}
where $0 \leq \lambda \leq 1$ controls the bias–variance trade-off.  

PPO updates the policy by maximizing the clipped surrogate objective, which prevents large, destabilizing changes:
\begin{equation}
L^{\mathrm{CLIP}}(\theta) = \mathbb{E}_t \!\left[ \min \big( r_t(\theta) \hat{A}_t, \ \text{clip}(r_t(\theta), 1-\epsilon, 1+\epsilon) \hat{A}_t \big) \right],
\label{eq:clip_objective}
\end{equation}
where $r_t(\theta) = \frac{\pi_\theta(\va_t|\vs_t)}{\pi_{\theta_\mathrm{old}}(\va_t|\vs_t)}$ is the probability ratio between the new and old policies, and $\epsilon$ is the clipping threshold.  Finally, the policy and value networks are updated by stochastic gradient ascent and descent, respectively:
\begin{align}
\theta &\leftarrow \theta + \alpha_\theta \nabla_\theta L^{\mathrm{CLIP}}(\theta), \label{eq:policy_update} \\
\phi &\leftarrow \phi - \alpha_\phi \nabla_\phi \tfrac{1}{2} \big( V_\phi(\vs_t) - R_t \big)^2,  \label{eq:value_update}
\end{align}
where $\alpha_\theta$ and $\alpha_\phi$ are the learning rates of the policy and value networks, and $R_t = \sum_{k=t}^T \vartheta^{k-t} r_k$ denotes the observed return.

\subsection{PPO-based Algorithm for Resource Allocation}
\setlength{\textfloatsep}{7pt}
\begin{algorithm}[t]
\small
\caption{\small Training PPO-based Algorithm}
\label{alg:PPO-FAS-FPA}
\LinesNumbered
\KwIn{Environment; initial policy $\theta$; value parameter $\phi$; learning rates $\alpha_\theta,\alpha_\phi$; clipping threshold $\epsilon$; discount factor $\vartheta$; GAE parameter $\lambda$.}
\KwOut{Trained policy $\theta^\star$ and value $\phi^\star$.}

\textbf{Initialization}: randomly initialize $\theta,\phi$\;

\For{each iteration}{
  \tcp{Environment interaction}
  \For{each time step $t$}{
    Observe state $\vs_t$ and sample action $\va_t=(\zeta_t,\kappa_t,\nu_t)$ from policy $\pi_\theta$\;
    Compute $I_{\mathrm{AP},m}$ via \eqref{eq:importance_score} and activate $\zeta_t \%$ APs with the largest $I_{\mathrm{AP},m}$\;
    Determine $\{N_m\}$ via \eqref{eq:antennas_allocation} with $\kappa_t$\;
    Allocate transmit power $\{\eta_{mk}\}$ via \eqref{eq:power_allocation} with $\nu_t$\;
    Compute reward $r_t$ via \eqref{eq:reward_function} and observe next state $\vs_{t+1}$\;
    Store transition $(\vs_t,\va_t,r_t,\vs_{t+1})$\;
  }

  \tcp{Policy update}
  Estimate $\hat A_t$ using GAE \eqref{eq:advantage_func}\;
  Update policy and value parameters based on \eqref{eq:policy_update}, \eqref{eq:value_update}
}
\end{algorithm}

Algorithm~\ref{alg:PPO-FAS-FPA} outlines the training process of the proposed PPO-based algorithm for antenna selection and power allocation. The agent interacts with the CFmMIMO environment by observing the current state $\vs_t$ and sampling an action $\va_t = (\zeta_t, \kappa_t, \nu_t)$ from the policy network $\pi_\theta$. These parameters configure the AP activation, antenna allocation, and power distribution mechanisms. Specifically, a subset of APs with the highest $I_{\mathrm{AP},m}$ is activated according to the AP activation ratio $\zeta_t$, antennas are allocated to active APs based on $\kappa_t$, and user power coefficients $\eta_{mk}$ are determined by $\nu_t$. After executing $\va_t$, the agent receives a reward $r_t$, and stores the transition $(\vs_t, \va_t, r_t, \vs_{t+1})$. Once a batch of transitions is collected, the policy and value networks are updated using the PPO objective with GAE. Through iterative interaction and policy refinement, the agent learns to map large-scale fading coefficients to the optimal values of $\zeta$, $\kappa$, and $\nu$. These learned coefficients are then applied in \eqref{eq:antennas_allocation} and \eqref{eq:power_allocation} to determine the number of antennas assigned to each AP and the power distribution among users.

\textbf{Complexity analysis: }The AP activation mechanism requires $\mathcal{O}(MK + M \log M)$ operations due to the computation and sorting of $I_{\mathrm{AP},m}$. After AP selection, the number of active APs reduces to $M_{\mathsf{active}} \le M$, lowering the cost of antenna selection to $\mathcal{O}(M_{\mathsf{active}})$. Power allocation requires computing coefficients for all active links, yielding complexity $\mathcal{O}(M_{\mathsf{active}} K)$, making the proposed approach suitable for real-time implementation.

\section{Performance Evaluation}
\label{sec:performance_evaluation}
\begin{figure*}[t!]
    \centering
    \begin{subfigure}[b]{0.32\textwidth}
        \centering
        \includegraphics[width=\textwidth]{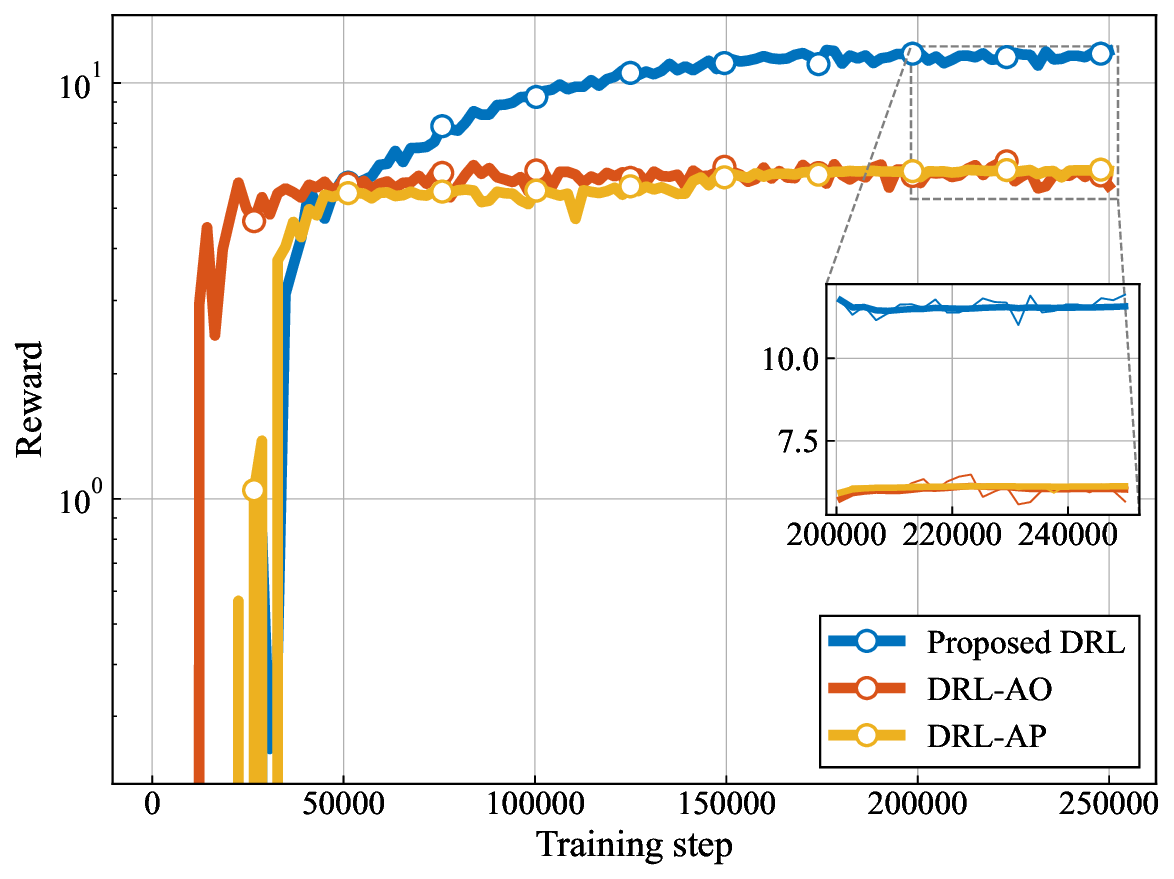}
        \caption{Convergence of the training reward.}
        \label{fig:convergence}
    \end{subfigure}
    \hfill
    \begin{subfigure}[b]{0.32\textwidth}
        \centering
        \includegraphics[width=\textwidth]{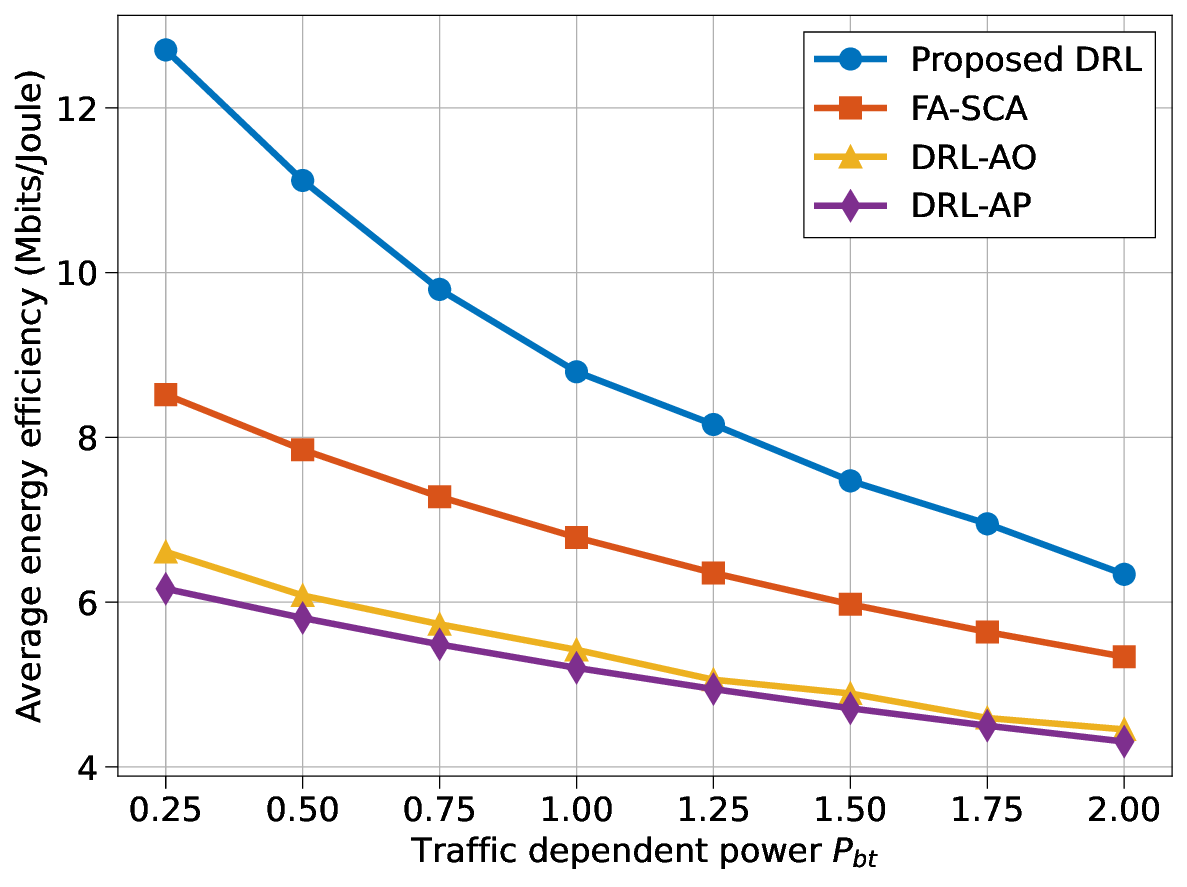}
        \caption{Energy efficiency versus traffic-dependent power $P_{bt}$.}
        \label{fig:ee_vs_pbt}
    \end{subfigure}
    \hfill
    \begin{subfigure}[b]{0.32\textwidth}
        \centering
        \includegraphics[width=\textwidth]{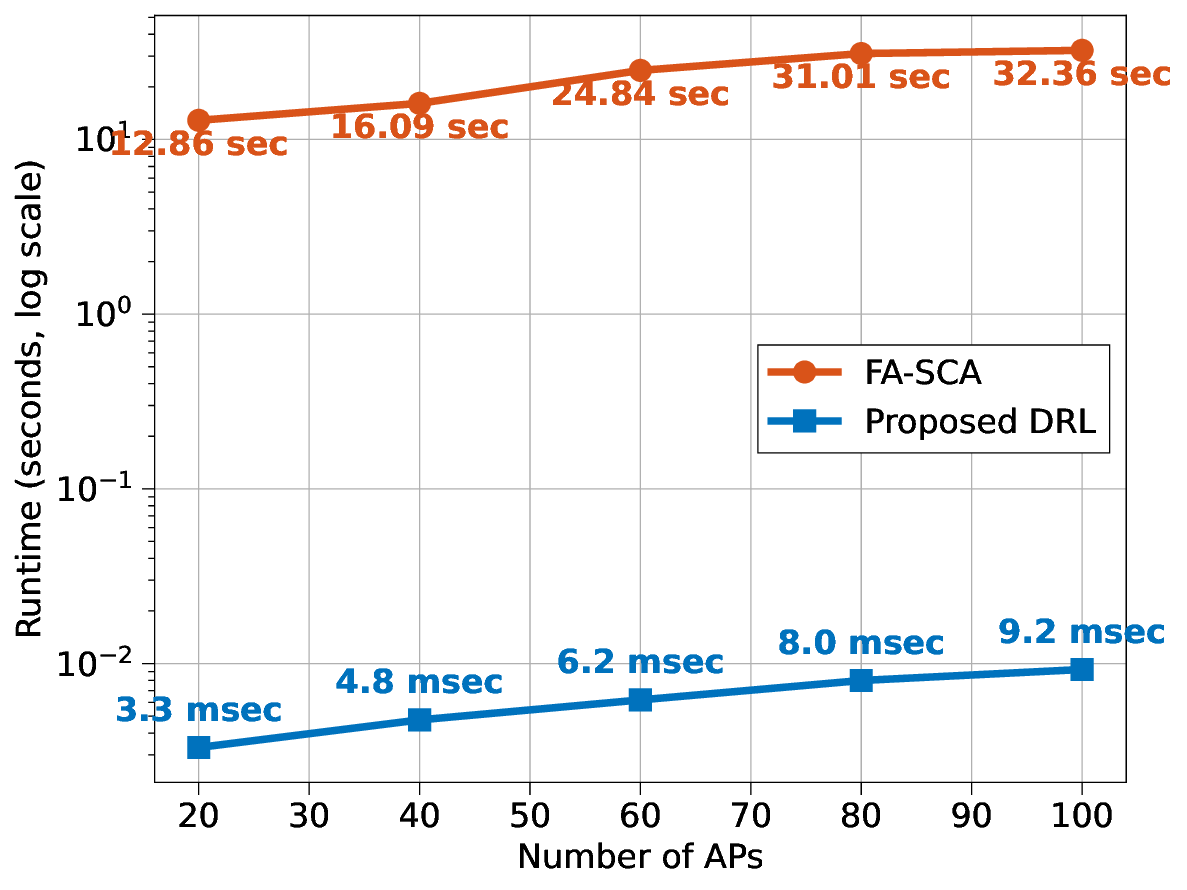}
        \caption{Average computation time versus the number of APs.}
        \label{fig:runtime}
    \end{subfigure}
    \caption{Convergence, energy efficiency, and runtime performance of the proposed DRL scheme compared with benchmarks.}
    \label{fig:all_results}
\end{figure*}
We consider a square network area of $1 \times 1$ km$^2$, wherein the APs and users are uniformly and independently distributed, with wrap-around at the edges to avoid boundary effects. The large-scale fading coefficient is modeled as $\beta_{mk} = \mathrm{PL}_{mk} \cdot z_{mk}$, where $z_{mk}$ is log-normal shadowing with standard deviation $\sigma_{\mathrm{sh}} = 8$~dB. The path loss $\mathrm{PL}_{mk}$ (in dB) follows the three-slope model \cite{ngo2017total}:
\begin{align*}
&\mathrm{PL}_{mk} = \\
&\begin{cases}
    -L - 35\log_{10}(d_{mk}), & \text{if } d_{mk} > d_1 \\
    -L - 15\log_{10}(d_1) - 20\log_{10}(d_{mk}), & \text{if } d_0 < d_{mk} \le d_1 \\
    -L - 15\log_{10}(d_1) - 20\log_{10}(d_0), & \text{if } d_{mk} \le d_0
\end{cases}
\end{align*}
where $d_{mk}$ is the distance between AP~$m$ and user~$k$ in meters, $d_0=10$ m, $d_1=50$ m, and $L=140.7$ dB. 
We consider the bandwidth $B = 20$~MHz and assume the receiver noise figure of $9$~dB. The power parameters are set as $\alpha_m = 0.4$, $P_{\mathrm{tc},m} = 0.2$~W, $P_{0,m} = 0.825$~W, $P_{\mathrm{bt},m} = 0.25$~W/(Gbit/s), $\rho_\mathsf{d} = 1$~W, and $\rho_\mathsf{p} = 0.2$~W. Furthermore, we set $\tau_\mathsf{p} = 20$, $\tau_\mathsf{c} = 200$ samples, and $S_{\mathrm{ok}} = 1$~bit/s/Hz. 

The DRL agent is trained using the PPO algorithm~\cite{schulman2017proximal}. Both the actor and critic networks consist of two hidden layers with 256 neurons each and ReLU activations. The learning rate and reward penalty coefficient are set to $3\times10^{-4}$ and $\xi_{\mathrm{pen}} = 20$, respectively, and the agent is trained for 300,000 timesteps using a minibatch size of 64. Other key hyperparameters of the PPO algorithm are the discount factor $\vartheta = 0.99$, the clipping parameter $\epsilon = 0.2$, and the GAE parameter $\lambda = 0.95$.

We compare the proposed DRL scheme against three benchmarks: The \textbf{DRL-AP} scheme assumes all APs are active ($\zeta=1$) and optimizes only antenna allocation ($\kappa$) and power control ($\nu$), isolating the effect of AP deactivation. The \textbf{DRL-AO} scheme optimizes only AP activation ($\zeta$), with $\kappa=0$ and $\nu=1$, highlighting the benefits of network sparsification. Finally, \textbf{FA-SCA} uses all antennas and optimizing power allocation using the SCA method \cite{ngo2017total}.

Fig.~\ref{fig:convergence} shows the convergence during training the proposed DRL, DRL-AO, and DRL-AP models with $M=40$, $K=20$, and $N=20$. All agents exhibit steadily increasing and stable reward curves, confirming the stability of the PPO-based algorithm. The proposed DRL framework achieves the highest reward, demonstrating that jointly optimizing AP deactivation ($\zeta$), antenna allocation ($\kappa$), and power control ($\nu$) is crucial for maximizing EE. The convergence speed reflects task complexity. The DRL-AO agent, which learns only $\zeta$, converges fastest after about 35,000 steps. 
In contrast, the proposed DRL framework, which must learn the interdependence among $\zeta$, $\kappa$, and $\nu$, stabilizes later, around 175,000 steps, while the DRL-AO agent converges after 50,000 steps.

Fig.~\ref{fig:ee_vs_pbt} compares the average EE under varying traffic-dependent power $P_{\mathrm{bt}}$ for a system with $M = 40$, $K = 20$, and $N = 20$. The proposed DRL-based framework consistently achieves the highest EE across all values of $P_{\mathrm{bt}}$, demonstrating the benefit of jointly optimizing AP activation, antenna allocation, and power control. 
For example, at $P_{\mathrm{bt}} = 0.25$ W/(Gbit/s), the proposed DRL achieves approximately 12.7 Mbits/Joule, yielding EE improvements of 50\% and 92\% compared to the FA-SCA scheme (8.5 Mbits/Joule) and the DRL-AO benchmark (6.6 Mbits/Joule), respectively.
The FA-SCA scheme, which keeps all APs and antennas active, achieves higher EE than DRL-AO, indicating that the SE gain from its optimal power allocation has a greater impact on EE than the energy saved by turning off inefficient APs.
Finally, the DRL-AP agent, which keeps all APs active and optimizes only $\kappa$ and $\nu$, achieves the lowest EE. This confirms that the energy savings from antenna deactivation alone are insufficient to compensate for its suboptimal power allocation. It highlights that without AP deactivation, the static circuit and backhaul power dominate, and the improvements from optimizing antennas and power cannot compensate for this inefficiency.

Finally, we evaluate the computational complexity, a key factor for practical deployment. Fig.~\ref{fig:runtime} compares the inference runtime of our DRL framework with the iterative FA-SCA benchmark for a network with $K = 20$ and $N = 20$. The FA-SCA method incurs prohibitively high latency, increasing from 12.86 seconds for $M = 20$ to 32.36 seconds for $M = 100$, due to repeatedly solving complex convex subproblems. In contrast, our DRL framework requires only a single forward pass and low-complexity resource allocation mechanisms, resulting in runtimes of just 3.3–9.2 ms-over 4,000 times faster at $M = 60$. These results demonstrate that the proposed method not only achieves higher EE but is also computationally efficient for real-time operation in large-scale cell-free networks.

\section{Conclusion}
\label{sec:conclusion}
In this paper, we proposed a novel joint antenna activation and power allocation to maximize EE in CFmMIMO systems. To address its high-dimensional, mixed-integer nature, we proposed a framework in which a DRL agent learns to control AP deactivation, antenna allocation, and power distribution across the network. Selected APs refine antenna selection and power distribution via computationally efficient resource allocation mechanisms, leveraging the decisions from the DRL agent. 
The simulation results demonstrated that our approach yields a more than twofold improvement in EE compared to FA-SCA, while drastically reducing runtime.

\section*{Acknowledgment}
This work was supported in part by European Union through MSCA Doctoral Network EXACT-6G (GA 101120297).
\bibliographystyle{IEEEtran}
\bibliography{bib/IEEEabrv,bib/Bibliography}
\end{document}